\documentclass[]{aa}

\usepackage{natbib}

\begin{document}

\title{The stellar content of the ring in NGC~660
\thanks{
Based on observations made with the NASA/ESA Hubble
Space Telescope, obtained from the Space Telescope Science Institute, which is operated by the
Association of Universities for Research in Astronomy, Inc., under NASA contract NAS5-26555.
}}
\author{G.M.\,Karataeva\inst{1,3}
\and N.A.\,Tikhonov\inst{2,4}
\and O.A.\,Galazutdinova\inst{2,4}
 \and V.A.\,Hagen-Thorn\inst{1,3}
\and V.A.\,Yakovleva\inst{1,3}}

\titlerunning{The Stellar Content of NGC~660}
\authorrunning{Karataeva et al.}

\offprints{G.\,Karataeva; \email{narka@astro.spbu.ru}}

\institute{Astronomical Institute, St.Petersburg State University,
Universitetsky pr., 28, Petrodvoretz, St.Petersburg, 198504, Russia
\and Special Astrophysical Observatory, N.Arkhyz, Karachai-Circassian
Rep., 369167, Russia
\and Isaac Newton Institute of Chile, St.Petersburg Branch
\and Isaac Newton Institute of Chile, SAO Branch}

\abstract{
We present the results of stellar photometry of the polar-ring galaxy
\object{NGC~660} using the Hubble Space Telescope's archival data obtained with
the Wide Field Planetary Camera 2. The final list of the resolved stars
contains 550 objects, a considerable part of which are blue and red
supergiants belonging to the polar ring. The analysis of the Colour Magnitude
Diagram for polar ring stars shows that it is best represented by
the isochrones with metallicity Z = 0.008. The process of star formation in the
polar ring was continuous and the age of the youngest detected stars is about 7~Myr.
\keywords{galaxies: individual: NGC 660 --- galaxies:
peculiar --- galaxies: starburst --- galaxies: stellar content}
}

\date{Received 15 January 2004 / Accepted 27 February 2004}

\maketitle

\section{Introduction}

NGC~660, known for a long time as a peculiar galaxy containing two
inclined dust lanes \citep{benv76}, was included by \citet{whit90} in their
catalogue of Polar-Ring Galaxies (PRGs), candidates and related
objects as a possible candidate (C\,-\,13). The existence
of two different kinematic systems was established from radio
H{\sc i} \citep{gott90} and CO \citep{comb92} observations, and the
galaxy was classified as a kinematically confirmed PRG \citep{arnab93}.

An extensive study of NGC~660 was performed by \citet{vanDr95}.
The authors give a detailed description of the host galaxy
(which is spiral, while host galaxies in PRGs are usually lenticular
or elliptical) and its ring, which is not polar but is inclined
to the disc of the host galaxy by $55\degr$. The authors  notice that 
the ring is very massive and  may be stable \citep{sparke86}. Multicolour
optical observations made with Schmidt telescope with low resolution 
($5\farcs5$) show that the ring is blue ($V-I_{\rm c}~=1\fm0$). 
From comparison with models of single burst star formation, the authors 
claim that the age of the ring is about 2~Gyr. However, many H{\sc ii} regions
\citep{young88, armus} observed in the ring  and the high mass of molecular
gas sufficient for active star formation \citep{comb92} may point to more
recent star formation. Detecting blue supergiants in the
ring may confirm this suggestion.

The galaxy is one of the nearest PRGs (D$\sim13$~ Mpc,
H$_{\rm 0}$~=75~km/s/Mpc) and resolving its ring into stars probably
may be achieved using HST images. Our study of PRGs NGC~2685 and NGC~4650A
whose distances are further than NGC~660 confirms this possibility
(Karataeva et al.,\,2004, Paper I). The HST archive contains several images 
of NGC~660,
unfortunately not very deep. These images form the observational basis for
our work devoted to resolving the ring of the galaxy into stars.

\begin{figure}[htb]
\vbox{\includegraphics{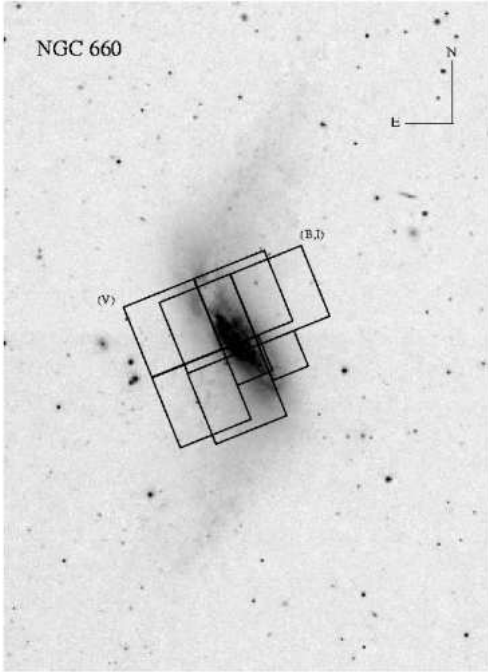}
\vspace{9.5cm}
}
\caption{DSS2 $8\arcmin \times 11\arcmin$ image of NGC~660 with WFPC2 footprint overlaid.
\label{f:N660_ima}
}
\end{figure}

\section{Observations and basic data processing}

We use archival HST/WFPC2 data of NGC 660 (see Table~\ref{t:Obs}). These observations were
carried out as part of Proposals N~9042 and N~5446. The data set consists of
images $F450W$, $F814W$, $F660W$-bands with total exposure times of 460~sec,
460~sec and 160~sec accordingly. Unfortunately, these images cover only a
part of the ring (see Fig.~\ref{f:N660_ima}). The total size of NGC~660 
according
to RC3 is $8\farcm3 \times 3\farcm2$. At the distance of NGC~660 the scale
is $\simeq$~65~pc in $1\arcsec$.

\begin{table}
\footnotesize
\caption{Observation log}
\renewcommand{\tabcolsep}{2pt}
\begin{tabular}{lcr}\\ \hline \hline
\multicolumn{1}{c}{Date}&
\multicolumn{1}{c}{Band}&
\multicolumn{1}{c}{Exposure}\\
                    \hline\\
  04.07.1995 & F660w&   $2\times80 sec$ \\
  03.07.2001 & F450w&   $2\times230 sec$ \\
             & F814w&   $2\times230 sec$\\

\hline
\end{tabular}
\label{t:Obs}
\end{table}

The raw frames were processed with the standard WFPC2 pipeline.
The data were extracted from the Archive using the On-The-Fly Reprocessing
(OTFR) STScI archive system, which reconstructs and calibrates the original 
data
with the latest calibration files, software and data parameters.
For details of the processing see Paper~I.

\begin{figure}[hbt]
\vbox{\includegraphics{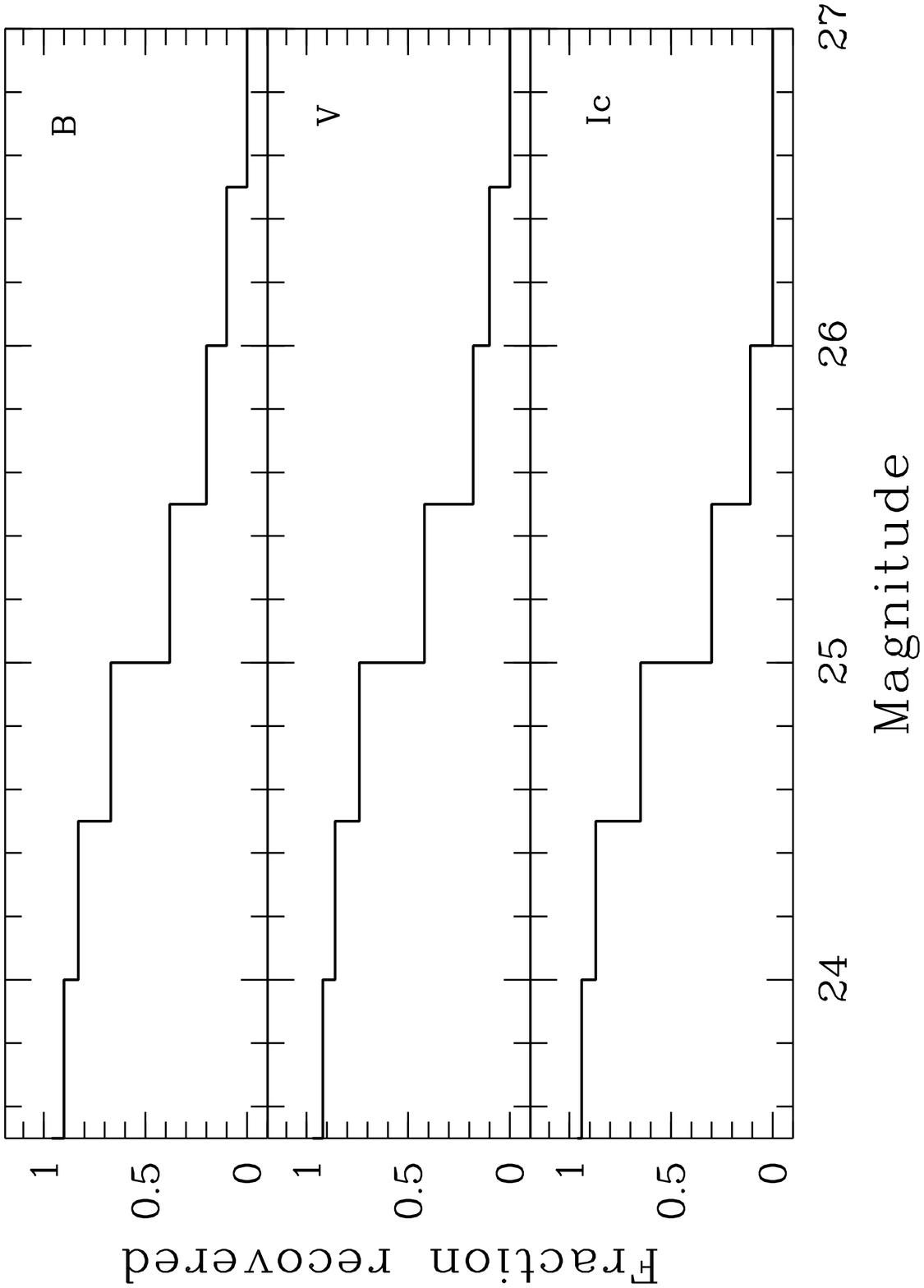}
     \includegraphics{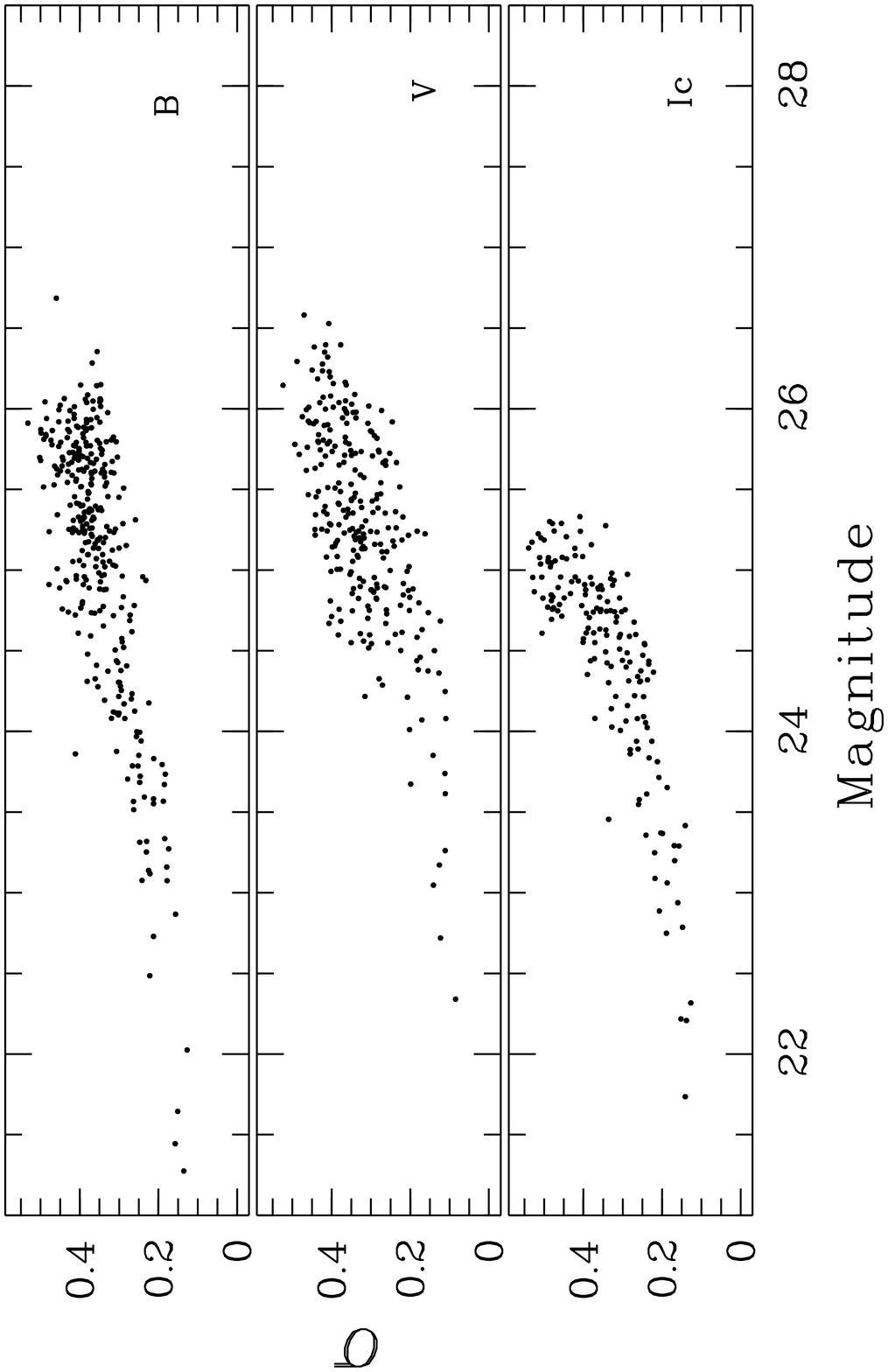}}
\vspace{7cm}
\caption{Completeness levels ({\em left}) and errors ({\em right})
of the WFPC2 photometry.
\label{f:Compl_err}
}
\end{figure}

The single-star photometry of the images was processed with
{\tt DAOPHOT}/{\tt ALLSTAR}. These programs use automatic star-finding
algorithms and then measure the stellar magnitude by fitting a PSF that is
constructed from the stars in uncrowded parts of the images.
Then we determined the aperture correction from the 1.5 pixel radius aperture
to the standard $0\farcs5$  radius  aperture size for the  WFPC2 photometric
system using bright uncrowded stars. The $F450W$, $F606W$ and $F814W$
instrumental
magnitudes were transformed to standard $BVI_{\rm c}$ in the Johnson-Cousins
system using the prescriptions of \citet{holtzman95}.
Background galaxies, unresolved blends and stars contaminated by
cosmetic CCD blemishes which have ``quality" parameters as defined in
{\tt ALLSTAR} in {\tt MIDAS} $|SHARP|>0.1, |CHI|>1.0$ were  eliminate from the
final list which contains 550 stars.

The results of the artificial star completeness test are shown in
Fig.~\ref{f:Compl_err},{\em left}. The recovery is
practically complete up to m~= $23\fm5$  in $B$-band , m~= $23\fm5$ in
$V$-band and m~= $24\fm0$  in $I_{\rm c}$-band.
Figure~\ref{f:Compl_err},{\em right} illustrates the precision of our
photometry.

\section{Results}

\begin{figure*}[hbt]
\vbox{\includegraphics{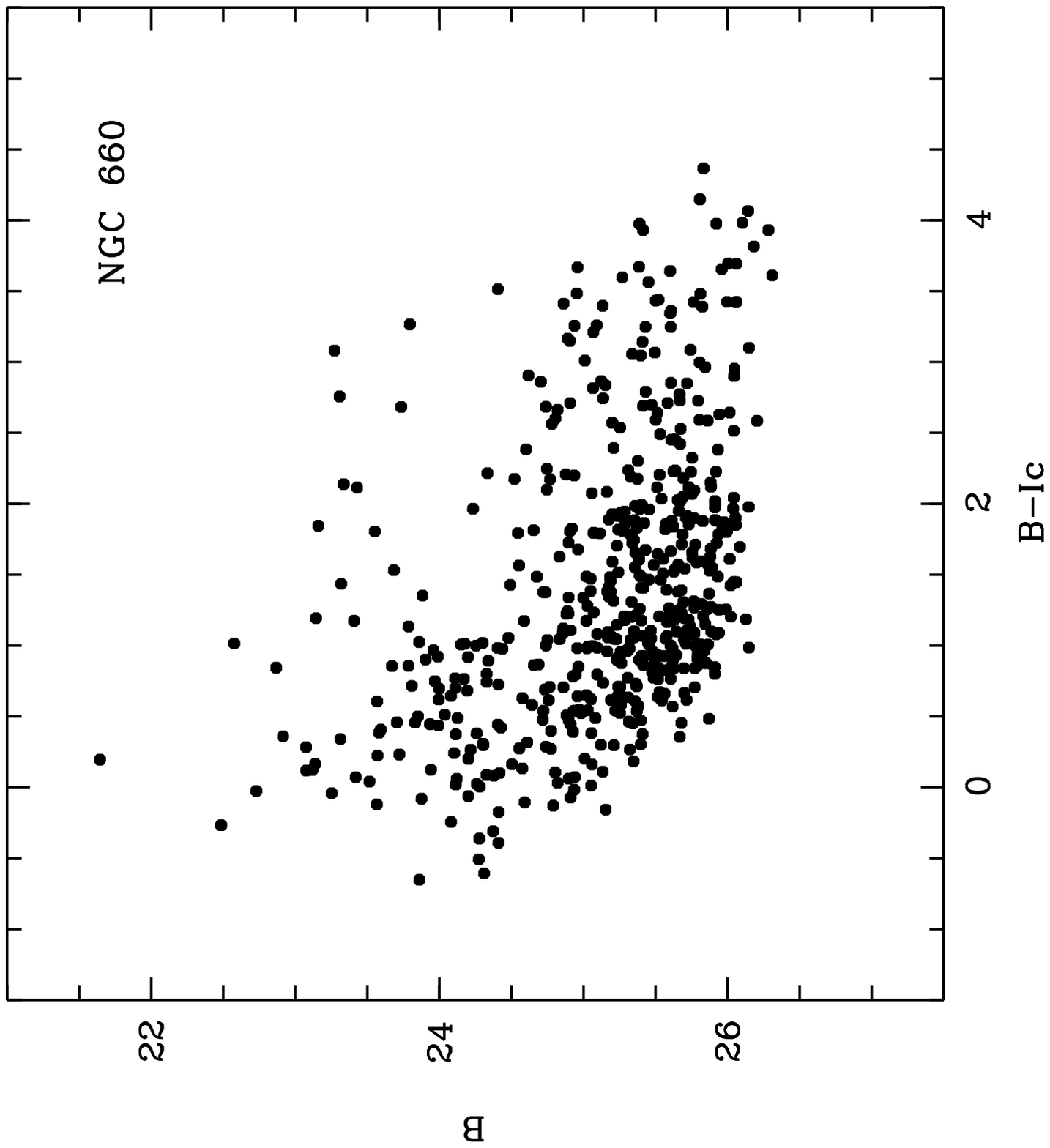}
      \includegraphics{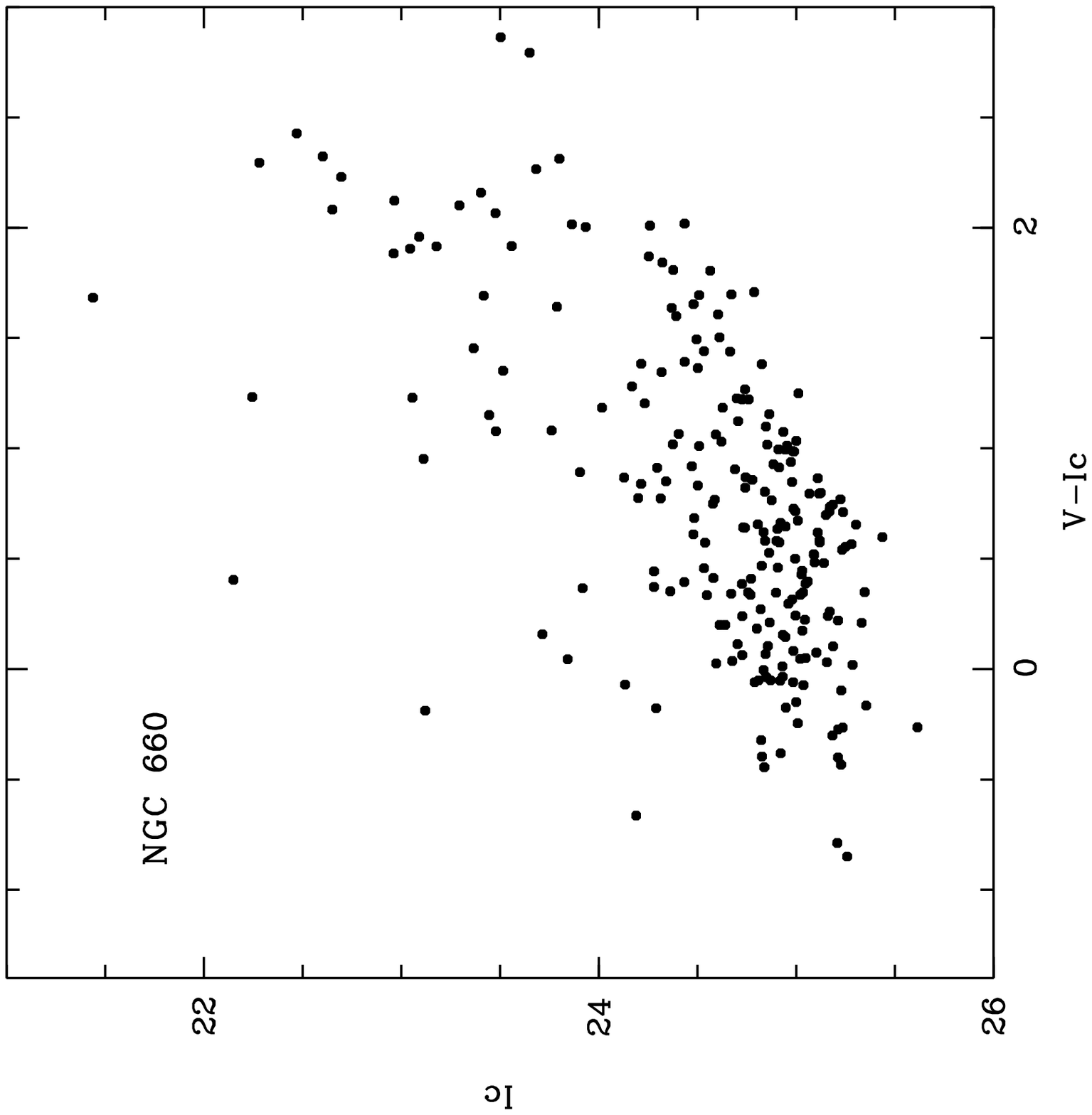}
\vspace*{7.5cm}
}
\caption{Colour--Magnitude Diagrams $B$ vs. $(B-I_{\rm c})$ ({\em left}) and
$I_{\rm c}$ vs. $(V-I_{\rm c})$ ({\em right}) corrected for
absorption in our Galaxy.
\label{f:CMD}
}
\end{figure*}

\begin{figure*}[htb]
\vbox{\includegraphics{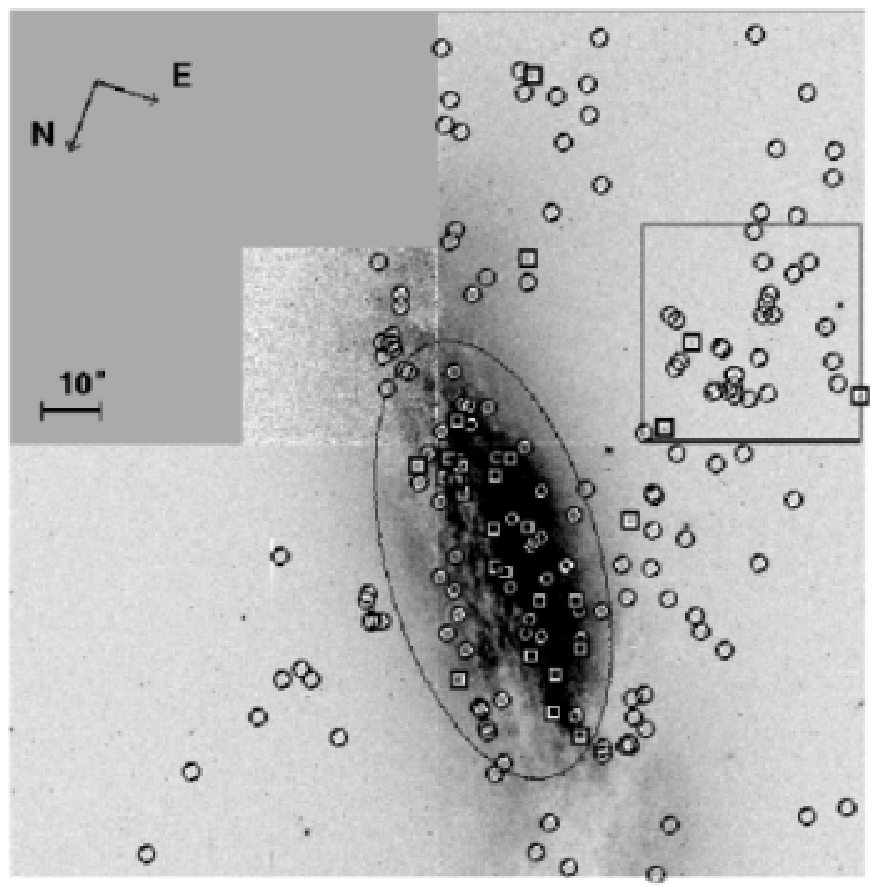}
      \includegraphics{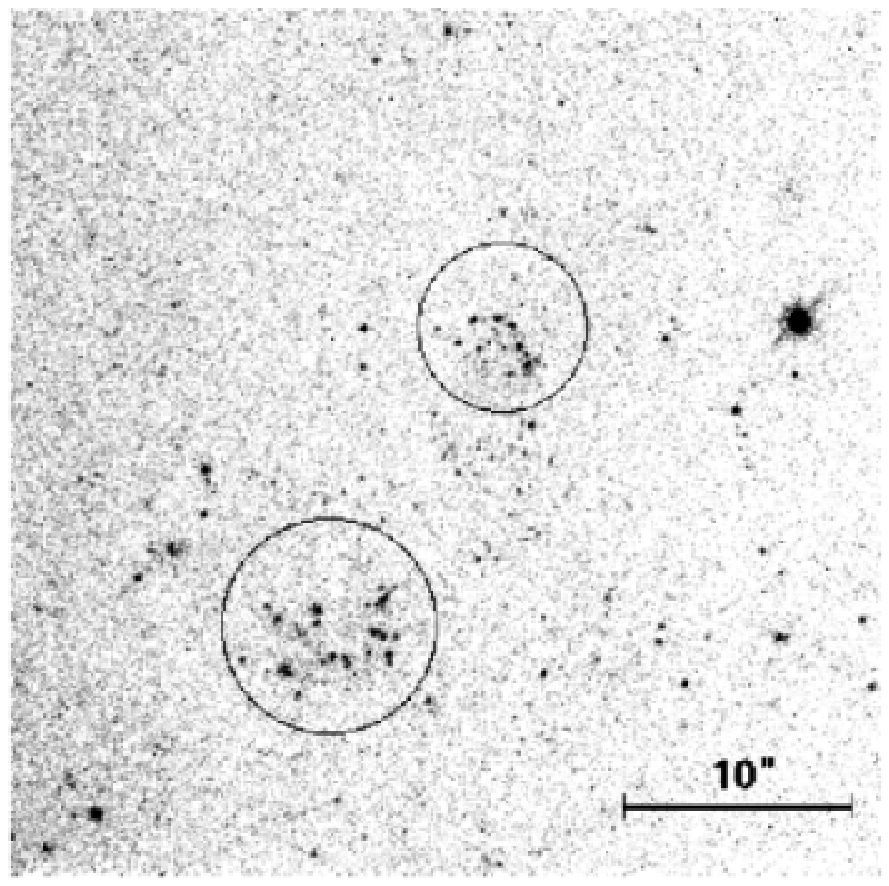}}
\vspace{8.6cm}
\caption{WFPC2 image of NGC~660.
{\em Left}: circles and squares indicate the position of blue and red 
supergiants, correspondingly. The stars located inside the ellipse
were rejected in construction of Fig.~\ref{f:CMD_iso}, right.
{\em Right}: two stellar complexes in the enlarged part (box on the 
{\em left}) of the NGC~660 polar ring.
\label{f:N660_ima2}
}
\end{figure*}

   The results of stellar photometry in the galaxy NGC~660
obtained after correcting for absorption in our Galaxy
($A_{\rm B} = 0\fm28$, $A_{\rm V} = 0\fm21$, $A_{\rm I_{\rm c}} =0\fm12$
\citep{schlegel}) are presented in Colour--Magnitude Diagrams (CMDs)
in Fig.~\ref{f:CMD}. For the CMD $B$~vs. $(B-I_{\rm c})$ there is a
concentration of blue stars in the region  $0 < (B-I_{\rm c}) < 1\fm0$
and $23\fm0 < B < 25\fm0$. Taking into account that there must be
intrinsic absorption in NGC~660 these stars may be considered as
blue supergiants by their colour indexes and luminosities (for NGC~660
the distance modulus is m-M~=$30\fm6$). The bulk of them are
located in the region of the polar ring (Fig.~\ref{f:N660_ima2}).

  While blue supergiants are detected in the CMD $B$ vs. $(B-I_{\rm c})$
quite reliably (the normal position for blue supergiants can be 
found from \citet{bertelli94}), the red supergiants, being near the 
limit in the
 $F450W$ image, are somewhat uncertainly. Nevertheless, some
 stars with $B < 25\fm5$ and $2\fm8 < (B-I_{\rm c})~< 4\fm0$
are located in the ring. The $F660W$ image is more favorable
for detecting red supergiants. Unfortunately, only a small part of the 
stars of the ring and host galaxy are represented in the CMD
 $I_{\rm c}$ vs. $(V-I_{\rm c})$ (Fig.~\ref{f:CMD}, right) because of
 the very short exposure time of the image as well as the large shift
between the $F660W$ and $F814W$ images. However, one can
see a concentration of stars in the region where red supergiants
may be located ($I_{\rm c}\approx22\fm5$ and $V-I_{\rm c}\approx2\fm0$).
A part of these stars are located in the ring.

\begin{figure*}[hbt]
\vbox{\includegraphics{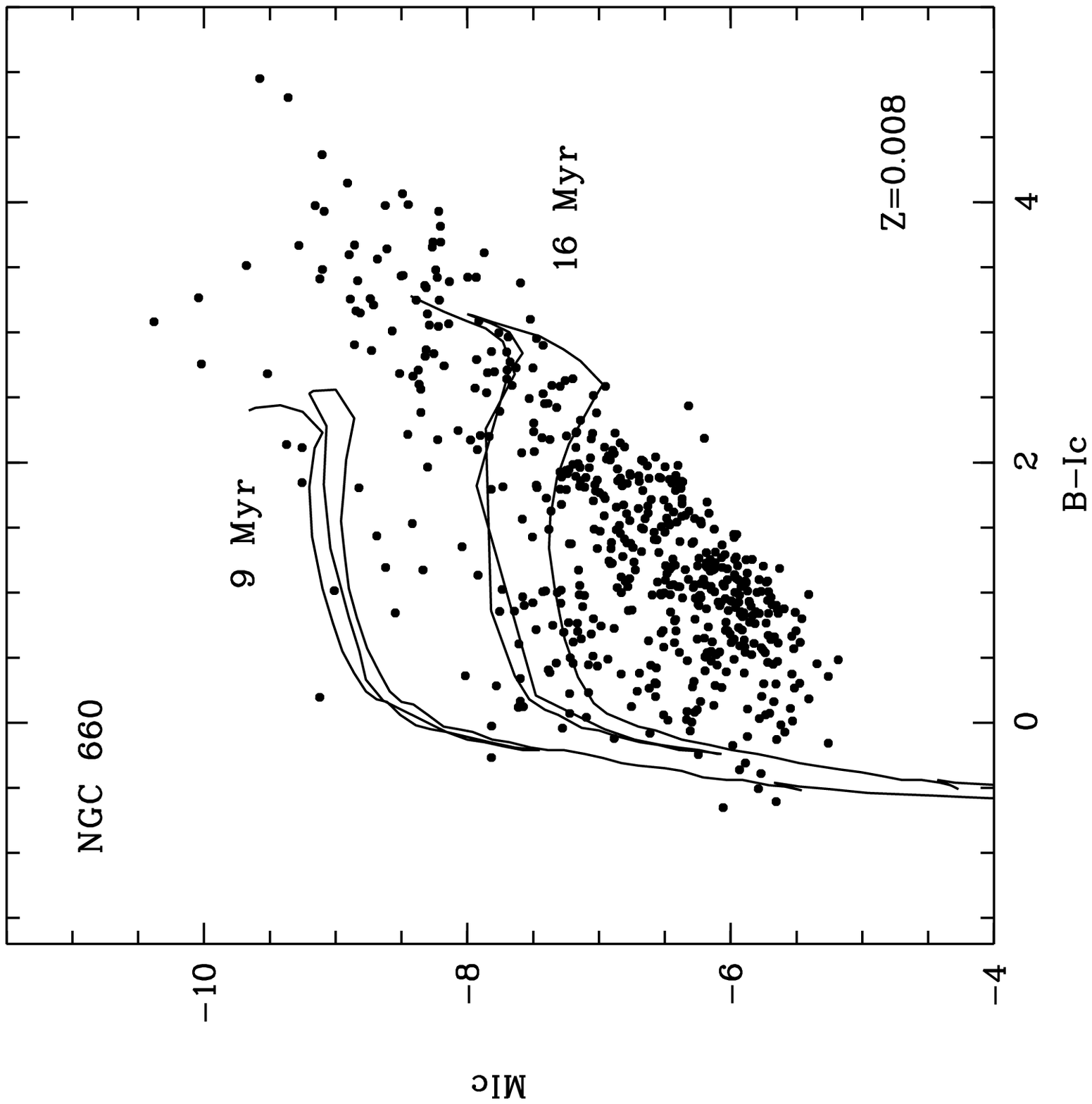}
      \includegraphics{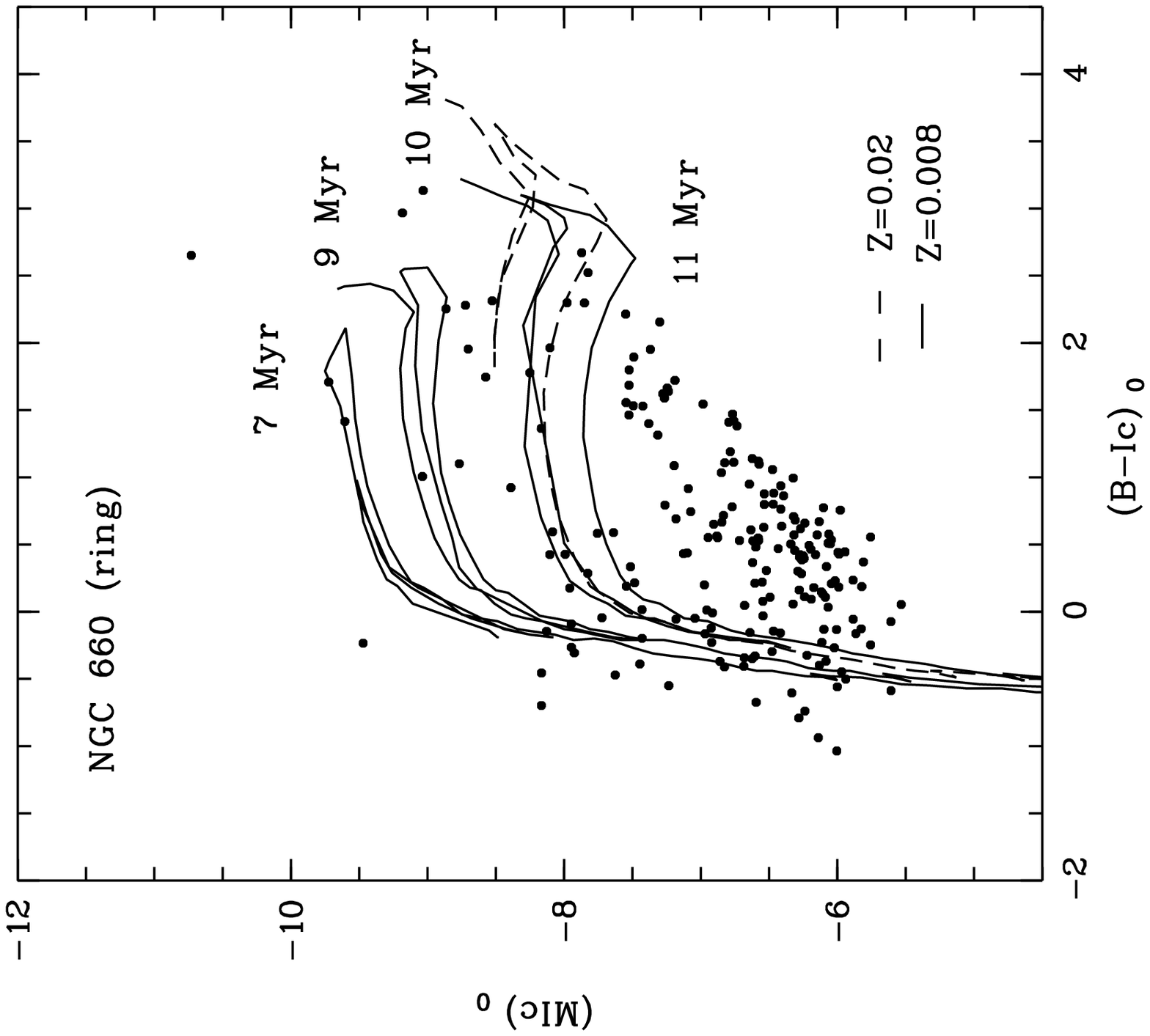}
\vspace*{7.5cm}
}
\caption{ M$_{\rm {I_{\rm c}}}$ vs $(B-I_{\rm c})$ CMDs of
NGC~660: {\em left} -- for all resolved stars; {\em right} -- for ring stars 
 after correction for intrinsic absorption in the ring. Stellar
 isochrones for the metallicity  Z=0.008 from the Padova library are overplotted.
 Dashed line on the {\em right} represents an example of a stellar
 isochrone for the metallicity  Z=0.02.
\label{f:CMD_iso}
}
\end{figure*}

 In Fig.~\ref{f:CMD_iso},{\em left} we give the observed CMD 
 M$_{\rm {I_{\rm c}}}$ vs $(B-I_{\rm c})$ for all resolved stars constructed
 with the distance modulus m-M~=$ 30\fm6$. Absolute magnitudes of
 the blue stars are normal for blue supergiants, and taking into
 account their positions (Fig.~\ref{f:N660_ima2}) we have no reason
 not to consider them as ring stars. This is confirmed, in particular,
 by the location of some stars in complexes, two of which are presented in
 Fig.~\ref{f:N660_ima2},{\em right}. One may expect that these complexes
 contain young stars. Such complexes were found in polar rings of the galaxies
 NGC~2685 and NGC~4650A (PaperI).

 The large width of the blue supergiants branch ($0 < (B-I_{\rm c}) < 1$)
 in Fig.~\ref{f:CMD},{\em left} and Fig.~\ref{f:CMD_iso},{\em left}
 may be explained by the different intrinsic absorption of  various stars.
 The presence of dust in the ring may be expected from the
 large amount of neutral hydrogen observed therein \citep{vanDr95}. The dust is
 seen directly as a dark strip in the region where the ring is projected
 on the disc of the galaxy. The dusty nature of the strip is confirmed
 by polarimetry \citep {resh91,alton}.

 In the latter paper it is found that the absorption of the light of galactic disc
 by the ring is equal to $1\fm2$ in the $V$- band (i.e. because
 A$_{\rm V}~=-2.5~lg~e^{\rm{-\tau_{\rm v}}}$ the optical depth of dust layers
 in the ring is $\tau_{\rm v}~=1.1$). This value gives the upper
 limit of absorption for the ring stars. Having no possibility to correct for
 absorption magnitudes and colour indexes of each star individually we correct
 CMDs for intrinsic absorption {\it on average}. If in the ring the dust and
 stars are mixed then for  $\tau_{\rm v}~=1.1$ one can find \citep{disney89}
 the mean absorption for the ring stars of $0\fm54$. For a normal reddening
 curve \citep{cardelli} this gives A$_{\rm B}~=0\fm69$, A$_{\rm I_{\rm c}}~=0\fm34$,
 E$_{\rm {(B-I_{\rm c})}}~=0\fm35$ and E$_{\rm {(V-I_{\rm c})}}~=0\fm22$. These 
 values may be used to correct CMDs.

 Because our goal is to investigate the stellar content of the ring we
 should eliminate the stars not belonging to it from the final diagram.
 The final CMD for polar ring stars after correcting for intrinsic absorption
 and eliminating stars of the host galaxy (made geometrically, as shown 
 in Fig.~\ref{f:N660_ima2},{\em left}) is
 given in Fig.~\ref{f:CMD_iso},{\em right}. Note that there is only a
 small difference between features in Figs.~\ref{f:CMD_iso},{\em right} and
 {\em left} in the region of blue stars, which,
 therefore, belong to the ring. Of course, the CMD may contain
 some contaminating objects from our Galaxy (for instance, the object
 which has $M_{\rm I_{\rm c}}\approx-10\fm8$ and $B-I_{\rm c}\approx2\fm6$),
 but their number is not large.

\section{Discussion}

The CMD for ring stars (Fig.~\ref{f:CMD_iso},{\em right}) may be used for
estimation of their ages and metallicity by comparison with calculated
isochrones. \citet{bertelli94} provide these isochrones for various 
metallicities. In
 Fig.~\ref{f:CMD_iso},{\em right} isochrones for the metallicity  Z=0.008
are overplotted on the CMD. Although the number of red supergiants is
small, they allow us to choose between the isochrones with different
 metallicity . The isochrones for the metallicity  Z=0.008 are in the
 best agreement with the observed CMD. The isochrones for higher metallicity
 (for example, solar, Z=0.02) contradict with the observed CMD (see 
 Fig.~\ref{f:CMD_iso},{\em right}) . A relatively low metallicity
 in the polar rings was found for NGC~2685 (PaperI) and some other
 polar rings \citep{radke03}. This is an important fact which should be
 taken into account in choosing a scenario of polar ring formation. 
 
 Fig.~\ref{f:CMD_iso},{\em right} shows that in the
 ring of NGC~660 there are very young stars with ages 7$\div$9~Myr.
Grouping the stars in complexes (see Fig.~\ref{f:N660_ima2},{right})
 shows that the stars are not yet dispersed in the stellar field, confirming
their youth.

 Because the CMD in Fig.~\ref{f:CMD_iso},{\em right} is not deep, we cannot
conclude anything about age of the ring. It may also contain old stars,
 like the Arp ring near M~81, where \citet{tikh04} detected
 red giants as well as young blue stars.

\begin{figure}[htb]
\vbox{\includegraphics{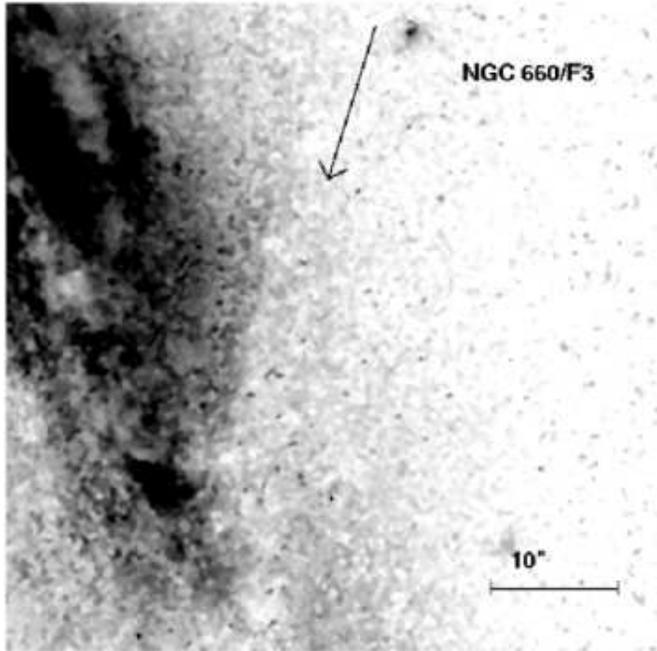}}

\vspace{8.6cm}
\caption{The distribution of the colour $B-I_{\rm c}$ in NGC~660 (chip WF3)
  (the scale is such that {\em dark} to {\em light} is the
  sense {\em red} to {\em blue}). The arrow indicates the direction of polar 
ring.
\label{f:N660_F3}
}
\end{figure}

 The relative youth of the ring is confirmed by its blue colour, as
 derived from surface photometry. \citet{vanDr95} showed that the ring
 with its colour index $V-I_{\rm c}\approx1\fm0$ is much bluer than the
 galaxy disc. Dividing frame $F450W$ by frame  $F814W$
 (see Fig.~\ref{f:N660_F3}) confirms this.
 By comparison of this colour index with models of galactic colour
 evolution, \citet{vanDr95} found that the age of the ring is about
 2$\times10^9$ years. However, the authors did not take into account
 any reddening. \citet{alton} corroborated this and found that after
 correcting for intrinsic absorption (which they somewhat
 overestimated) the colour-index reduces to $V-I_{\rm c}\approx0\fm7$ and
 age -- to $\sim10^8$years 
 (this value was found after an incorrect comparison of the colour index
 $V-I_{\rm c}$ (Cousins) with models, calculated for $V-I$ (Johnson)).
 Indeed, the full (Galactic and intrinsic) colour excess is
 E$_{\rm{V-I_{\rm c}}}~=0\fm31$ (see above) and correct comparison with
 the models calculated for $V-I_{\rm c}$ (Cousins) \citep{kurth99}
 give the age of the ring as 1$\times10^9$ years.

 A final conclusion about the age of the ring may be obtained
 only after constructing the CMD sufficiently deep to reach the region
 of red giants. 
 Now we can conclude only that very young stars exist in the ring
 and that the starforming process is continuous because in
 Fig.~\ref{f:CMD_iso},{\em right} there are many stars below the
 isochrones.

\section{Conclusions}

We have resolved the ring of NGC~660 into stars.
 Some blue and red supergiants were
found with ages of about 7$\div$9~Myr. The relatively
low metallicity of these stars is in accordance with the
data for polar rings in other PRGs.
        
It is very important to reach the region of red
giants in CMDs. This might give not only the information about
ring age but also assist in choosing the model of ring
formation because the merging scenario \citep{bourncomb03}
predicts the existence around the galaxy of a halo from
old/intermediate age stars. Longer exposures with HST
may allow us to construct such CMDs.

\begin{acknowledgements}

 This work was supported by RFBR via grants 03-02-16344 and 02-02-16033.
\end{acknowledgements}

\end{document}